\documentclass[manuscript]{aastex}
\usepackage{amsmath}
\usepackage{apjfonts}
\usepackage{graphicx}
\usepackage{natbib}

\newcommand{\kep}{$K_{ep}$}
\newcommand{\nh}{\bar{n}_{\rm H}}

\slugcomment{Accepted by ApJ. : v6.4}

\shorttitle{\emph{Fermi} LAT Observations of the SNR G8.7$-$0.1}

\shortauthors{\emph{Fermi}-LAT Collaboration}

\begin{document}

\title{\emph{Fermi} Large Area Telescope Observations of the Supernova Remnant G8.7$-$0.1}



\author{
M.~Ajello\altaffilmark{2}, 
A.~Allafort\altaffilmark{2}, 
L.~Baldini\altaffilmark{3}, 
J.~Ballet\altaffilmark{4}, 
G.~Barbiellini\altaffilmark{5,6}, 
D.~Bastieri\altaffilmark{7,8}, 
K.~Bechtol\altaffilmark{2}, 
R.~Bellazzini\altaffilmark{3}, 
B.~Berenji\altaffilmark{2}, 
R.~D.~Blandford\altaffilmark{2}
% ...
}
% ...
\altaffiltext{2}{W. W. Hansen Experimental Physics Laboratory, Kavli Institute for Particle Astrophysics and Cosmology, Department of Physics and SLAC National Accelerator Laboratory, Stanford University, Stanford, CA 94305, USA}
\altaffiltext{3}{Istituto Nazionale di Fisica Nucleare, Sezione di Pisa, I-56127 Pisa, Italy}
\altaffiltext{4}{Laboratoire AIM, CEA-IRFU/CNRS/Universit\'e Paris Diderot, Service d'Astrophysique, CEA Saclay, 91191 Gif sur Yvette, France}
\altaffiltext{5}{Istituto Nazionale di Fisica Nucleare, Sezione di Trieste, I-34127 Trieste, Italy}
\altaffiltext{6}{Dipartimento di Fisica, Universit\`a di Trieste, I-34127 Trieste, Italy}
\altaffiltext{7}{Istituto Nazionale di Fisica Nucleare, Sezione di Padova, I-35131 Padova, Italy}
\altaffiltext{8}{Dipartimento di Fisica ``G. Galilei", Universit\`a di Padova, I-35131 Padova, Italy}
% ...


\begin{abstract}
 We present a detailed analysis of the GeV gamma-ray emission toward
 the supernova remnant~(SNR) G8.7$-$0.1 with the Large Area
 Telescope~(LAT) onboard the \emph{Fermi} Gamma-ray Space
 Telescope. 
An investigation of the relationship among G8.7$-$0.1 and the TeV
 unidentified source HESS~J1804$-$216 provides us with an important clue
 on diffusion process of cosmic rays if particle acceleration operates
 in the SNR.
The GeV gamma-ray emission is extended with most of the emission in
 positional coincidence with the SNR G8.7$-$0.1 and a lesser part
 located outside the western boundary of G8.7$-$0.1.
The region of the gamma-ray emission overlaps spatially-connected
 molecular clouds, implying a physical connection for the gamma-ray structure.
The total gamma-ray spectrum measured with LAT from 200 MeV--100 GeV
can be described by a broken power-law function with a break of 2.4
 $\pm$ 0.6 (stat) $\pm$ 1.2 (sys) GeV, and photon indices of
 2.10 $\pm$ 0.06 (stat) $\pm$ 0.10 (sys) below the break and 2.70 $\pm$
 0.12 (stat) $\pm$ 0.14 (sys) above the break.
Given the spatial association among the gamma rays, the radio emission
 of G8.7$-$0.1, and the molecular clouds, 
the decay of $\pi^{0}$s produced by particles accelerated in the SNR
 and hitting the molecular clouds naturally explains the GeV gamma-ray
 spectrum.
We also find that the GeV morphology is not well represented by
 the TeV emission from HESS~J1804$-$216
and that the spectrum in the GeV band is not consistent with
 the extrapolation of the TeV gamma-ray spectrum. 
The spectral index of the TeV emission is consistent with the
particle spectral index predicted by a theory that assumes
energy-dependent diffusion of particles accelerated in an SNR. 
We discuss the possibility that the
 TeV spectrum originates from the interaction of
particles accelerated in G8.7$-$0.1 with molecular clouds, and
 we constrain the diffusion coefficient of the particles.

\end{abstract}


\section{Introduction}

Galactic cosmic rays are widely believed to be accelerated through the
diffusive shock acceleration process at the shock of supernova remnants
(SNRs)~\citep[and references therein]{Reynolds2008}.
It is generally expected that if a dense molecular cloud is overtaken by
a supernova blast wave, the molecular cloud can be illuminated
by relativistic particles accelerated at SNR shocks~\citep[e.g.][]{Aharonian94}.
 If the accelerated particles are comprised mostly of protons,
say $>100$ times more abundant than electrons like the
observed Galactic cosmic rays, decays of neutral pions produced in inelastic collisions
of the accelerated protons with dense gas are expected to be
a dominant radiation component in the gamma-ray spectrum of the
cosmic-ray-illuminated molecular cloud.
Thus, gamma-ray observations of SNRs interacting with adjacent molecular clouds are important for the study of cosmic rays.

The Large Area Telescope (LAT)
onboard the \emph {Fermi} Gamma-ray Space Telescope has recently detected
GeV gamma rays from several middle-aged SNRs interacting with molecular
clouds~\citep{W51C,W44,IC443,W28,W49B}.
The GeV emission from these SNRs is bright and spatially coincident with
molecular clouds, suggesting a hadronic origin as the most plausible explanation~\citep{W51C,W44,IC443,W28,W49B}.
In addition, the LAT spectra of these sources exhibit spectral breaks above
a few GeV and steepening above the breaks.
A possible conventional mechanism for these spectral properties
is the energy-dependent diffusion of accelerated particles from the SNR shell into nearby molecular
clouds~\citep[e.g.,][]{Aharonian96,Gabici07,Ohira11}.
On the other hand, \cite{Uchiyama10} indicated that reaccelerated
pre-existing cosmic-rays compressed at a radiative shock in a molecular
cloud can explain the flat radio spectra and high gamma-ray luminosity
observed in these SNRs and that the Alfv$\acute{\rm e}$n wave evanescence due to the
strong ion-neutral collisions at the shock can cause the spectral breaks.
Thus, the observation of GeV gamma rays from an additional SNR in this class adds valuable information for the study of cosmic-ray
acceleration in SNRs and their interactions with surrounding matter and/or magnetic
fields.

G8.7$-$0.1 is a middle-aged SNR located within W30~\citep{Ojeda02}, a massive star
forming region, and having nine discrete 
\ion{H}{2} regions along the southern boundary~\citep{Blitz82}.
In the radio band, the shell-like synchrotron emission has a diameter of
$\sim$ 45$'$ and a spectral index of $\alpha$~=~0.5~\citep{Kassim90},
suggesting that electrons are accelerated via diffusive
shock acceleration.
The conjunction of the molecular clouds associated with G8.7$-$0.1~\citep{Blitz82}
and an OH maser on the eastern edge of
the remnant~\citep{Hewitt09} imply that the SNR is interacting with those molecular
clouds. 
The northern part of the remnant is filled by a thermal X-ray plasma
observed by \emph{ROSAT}~\citep{Finley94}. The distance to
G8.7$-$0.1 is estimated to be
$\sim$ 4.8--6 kpc based on kinematic distances to the \ion{H}{2}
regions associated with the SNR~\citep{Kassim90, Brand93} and
3.2--4.3 kpc based of the SNR evolution with the observed X-ray
temperature and the angular radius~\citep{Finley94}.
The age of the SNR is estimated to be 1.5--2.8 $\times$ 10$^{4}$~yr based
on applying a Sedov solution to the X-ray observation under the assumption of
an initial kinetic energy of 10$^{51}$~erg~\citep{Finley94}, or alternatively,
1.5~$\times$~10$^{4}$~yr using the relation between the age and the
surface brightness~\citep{Odegard86}. 
In this paper, we adopt an age of 2.5~$\times$~10$^{4}$~yr.

The HESS collaboration found a TeV gamma-ray
source in the vicinity of G8.7$-$0.1, HESS~J1804$-$216, which has an
extension of 22$'$~\citep{HESSsurvey06} and has been confirmed by
CANGAROO-III~\citep{Higashi08}. This source lacks an evident counterpart and is classified as
unidentified. \cite{Gabici07} predicts that a number of TeV unidentified
sources might be explained by molecular clouds illuminated by cosmic
rays escaping from a nearby SNR.
Thus, the relationship between HESS~J1804$-$216 and G8.7$-$0.1 is interesting
for probing the diffusion process of cosmic rays assuming that G8.7$-$0.1 is a probable cosmic-ray accelerator.
Measurements with the Energetic Gamma-Ray Experiment Telescope (EGRET)
onboard the \emph{Compton Gamma-ray Observatory} found no gamma-ray
sources around G8.7$-$0.1. 
A gamma-ray source is listed around G8.7$-$0.1 in
the Astro-rivelatore Gamma a Immagini LEggero~(AGILE) one year
catalog~\citep{AGILEcatalog}. However, AGILE has not published a detailed analysis of this field.
Three LAT sources in the vicinity of G8.7$-$0.1 are listed in the 1FGL
catalog~\citep{1yrCatalog}, which is compiled under the assumption that
sources are point-like. 
1FGL J1805.2$-$2137c was studied by \cite{Castro10} as the SNR with a
point-like assumption.

In this paper, we report a detailed analysis of the LAT sources
around G8.7$-$0.1 based on 23-months data.
First, we give a brief description of the gamma-ray selection in
Section~\ref{sec:obs}. The analysis procedures and results are explained
in Section~\ref{sec:ana},
including measurements of the spatial extension and spectra of the LAT sources near
the remnant. The discussion is given in Section~\ref{sec:discuss}, followed
by conclusions in Section~\ref{sec:conclusion}.

\section{OBSERVATION \& DATA REDUCTION}

\label{sec:obs}
The LAT is the main instrument on \emph{Fermi}.
The energy band extends from $\sim$~20~MeV to $>$~300~GeV\footnote{As noted
below in the present analysis we use only events with energies $>$ 200~MeV.}.
It is an electron-positron pair production telescope consisting of layers of tungsten foils and silicon microstrip detectors to measure the arrival directions of incoming gamma rays, and a hodoscopic cesium iodide calorimeter to determine the gamma-ray energies.
The instrument is surrounded by 89 plastic scintillator tiles that serve as an anticoincidence detector for rejecting charged particle events.
Details of the LAT instrument and pre-launch expectations of the performance can be found in \cite{Atwood09}. 
Relative to earlier gamma-ray missions, the LAT has a large $\sim$~2.4~sr field of view, a large effective
area ($\sim$~8000~cm$^2$ for $>$1~GeV if on-axis), and improved angular resolution with a point spread function~(PSF) described by a 68\% containment angle better than 1$^\circ$ at 1~GeV.

Routine science operations of the LAT began on August 4, 2008, after the conclusion of a commissioning period.
We have analyzed events around G8.7$-$0.1 collected from August
4, 2008, to July 9, 2010, with a total exposure of $\sim$~5.5~$\times$~10$^{10}$~cm$^2$~s~(at 1~GeV).
During this time interval, the LAT was operating in sky survey mode nearly all of the time, obtaining complete sky coverage every 2 orbits~($\sim$~3~hours) and relatively uniform exposure over time.

We used the standard LAT analysis software, \emph{ScienceTools} v9r15,
which is available from the \emph{Fermi} Science Support Center (FSSC)\footnote{Software and documentation of the \emph{Fermi} \emph{ScienceTools} are distributed by \emph{Fermi} Science Support Center at http://fermi.gsfc.nasa.gov/ssc}, and applied the following event selection criteria: 
1) events should have the highest probability of being gamma rays, i.e., they should be classified as so-called \emph{Diffuse} class \citep{Atwood09}, 
2) the reconstructed zenith angles of the arrival direction of gamma rays should be less than 105$^\circ$ 
to minimize contamination by gamma rays from the limb of the Earth,
3) the center of the LAT field of view should be within 52$^\circ$ of zenith in order to exclude data from the short time intervals when the field of view contains a larger portion of the Earth.
No gamma-ray bursts were detected by the LAT within
15$^\circ$ of G8.7$-$0.1; thus, we did not need to
apply any additional time cut.
The energy range analyzed here is restricted to $>$~200~MeV to avoid possible large systematic uncertainties at lower energies due to the strong Galactic diffuse emission~(especially for G8.7$-$0.1, which lies in the direction of the Galactic center), 
smaller effective area, and broader PSF.

\section{ANALYSIS AND RESULTS}
\label{sec:ana}
\subsection{Morphological Analysis}
\label{subsec:extension}
\begin{figure}
\epsscale{1.10}
\plotone{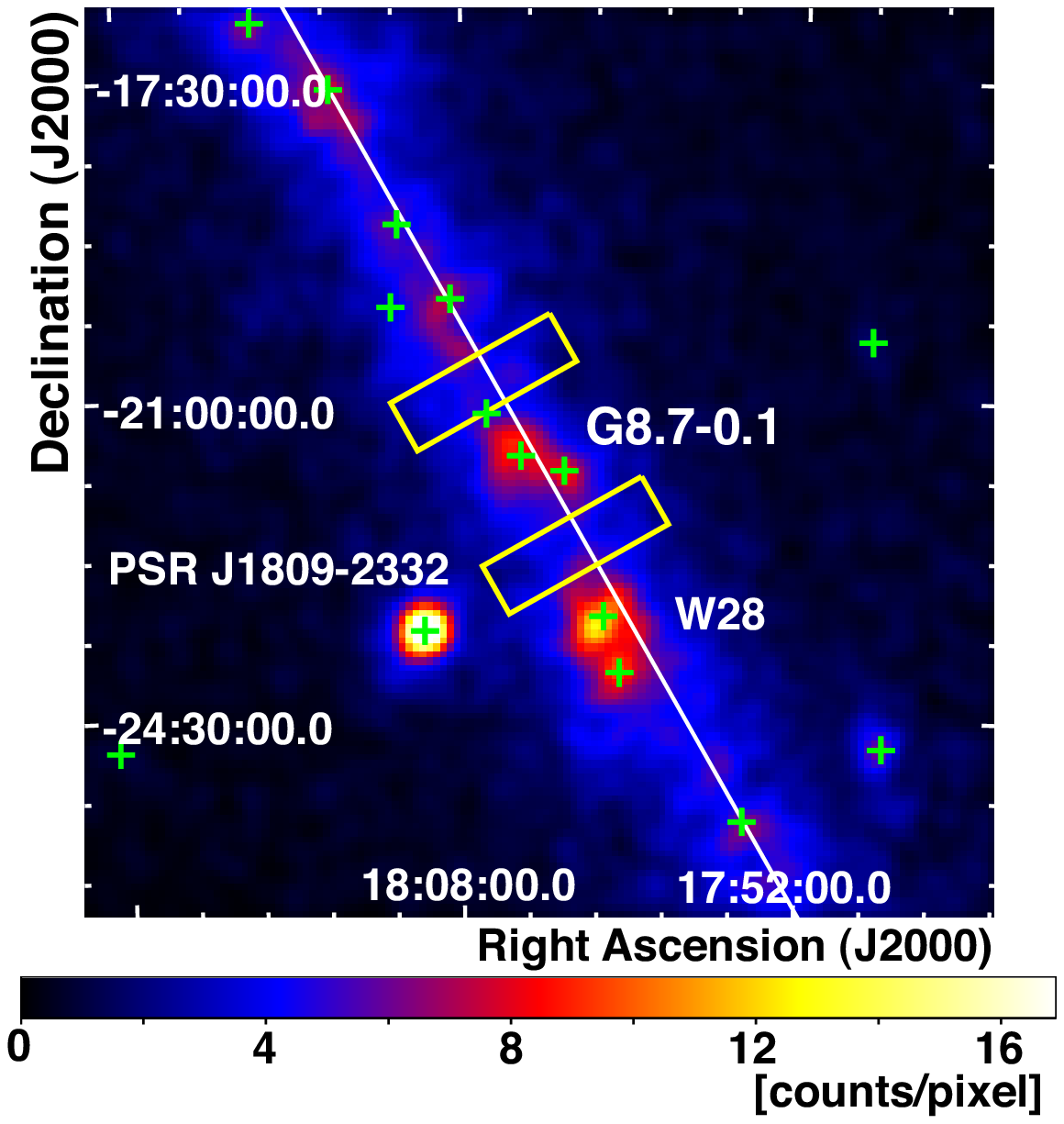}
\caption{\emph{Fermi} LAT 2-10~GeV counts map around the SNR G8.7-0.1. The count
 map is smoothed by a Gaussian kernel of $\sigma$~=~0$\fdg$225, with
 the pixel size of 0$\fdg$075. 
The green pluses indicate the sources in the 1FGL catalog~\citep{1yrCatalog}.
The yellow
 boxes indicate the regions that are used to evaluate the energy
 dependence for the systematic errors of the Galactic diffuse model.
The white line from top left to bottom right indicates the Galactic plane.}
\label{fig:cmap_wide}
\end{figure}

Figure~\ref{fig:cmap_wide} shows a smoothed count map in the
2--10~GeV energy band in a 10$^\circ~\times$~10$^\circ$ region around
G8.7$-$0.1. The average surface brightness of this region is about 2
times larger than neighboring regions along the Galactic plane.
There are three LAT sources in the vicinity of G8.7$-$0.1 in the 1FGL
catalog~\citep{1yrCatalog}: 1FGL J1805.2$-$2137c and 1FGL J1806.8$-$2109c
located to the east, and 1FGL J1803.1$-$2147c located to the west.

In order to evaluate the source extension and location of
these three sources, we used the maximum likelihood tool, {\tt gtlike},
which is available as part of the \emph{Fermi}
\emph{ScienceTools}. 
The likelihood is the product of the probabilities of observing the gamma-ray counts within each spatial and energy bin for a specified emission model. The best parameter values are estimated by maximizing the likelihood of the data describing the given model \citep{Mattox96}.
The probability density function for the likelihood analysis includes 
1)~individual sources detected in the 1FGL catalog within 15$^\circ$ of
G8.7$-$0.1,
2)~the Galactic diffuse emission resulting from cosmic-ray interactions with interstellar medium and radiation based on the LAT standard diffuse background model \emph{gll\_iem\_v02} available from the FSSC\footnote{The model can be downloaded from\\ http://fermi.gsfc.nasa.gov/ssc/data/access/lat/BackgroundModels.html.},
and 3)~an isotropic component to represent extragalactic gamma rays and residual
instrumental backgrounds using the standard isotropic spectral template \emph{isotropic\_iem\_v02} also available from the FSSC. 
The region of interest for the binned maximum likelihood analysis
was a square region of 20$^\circ~\times$~20$^\circ$
in Galactic coordinates centered on G8.7$-$0.1 with a pixel size of 0$\fdg$1.
The instrument response functions~(IRFs) used in our work were the ``Pass~6~v3~Diffuse'' (P6\_V3\_DIFFUSE) IRFs; a post-launch update to address gamma-ray detection inefficiencies that are correlated with background rates. 
Since PSR~J1809$-$2332~\citep{PulsarCatalog} and W28~\citep{W28} are
the brightest sources in the region of interest, they must be carefully
modeled to perform the morphological studies.
Their spectral shapes were modeled as a
power$-$law with an exponential cutoff and a broken power$-$law,
respectively. A spatial template was used for W28 to take into account its extension~\citep{W28}.

\begin{figure}
 \includegraphics[scale=.45]{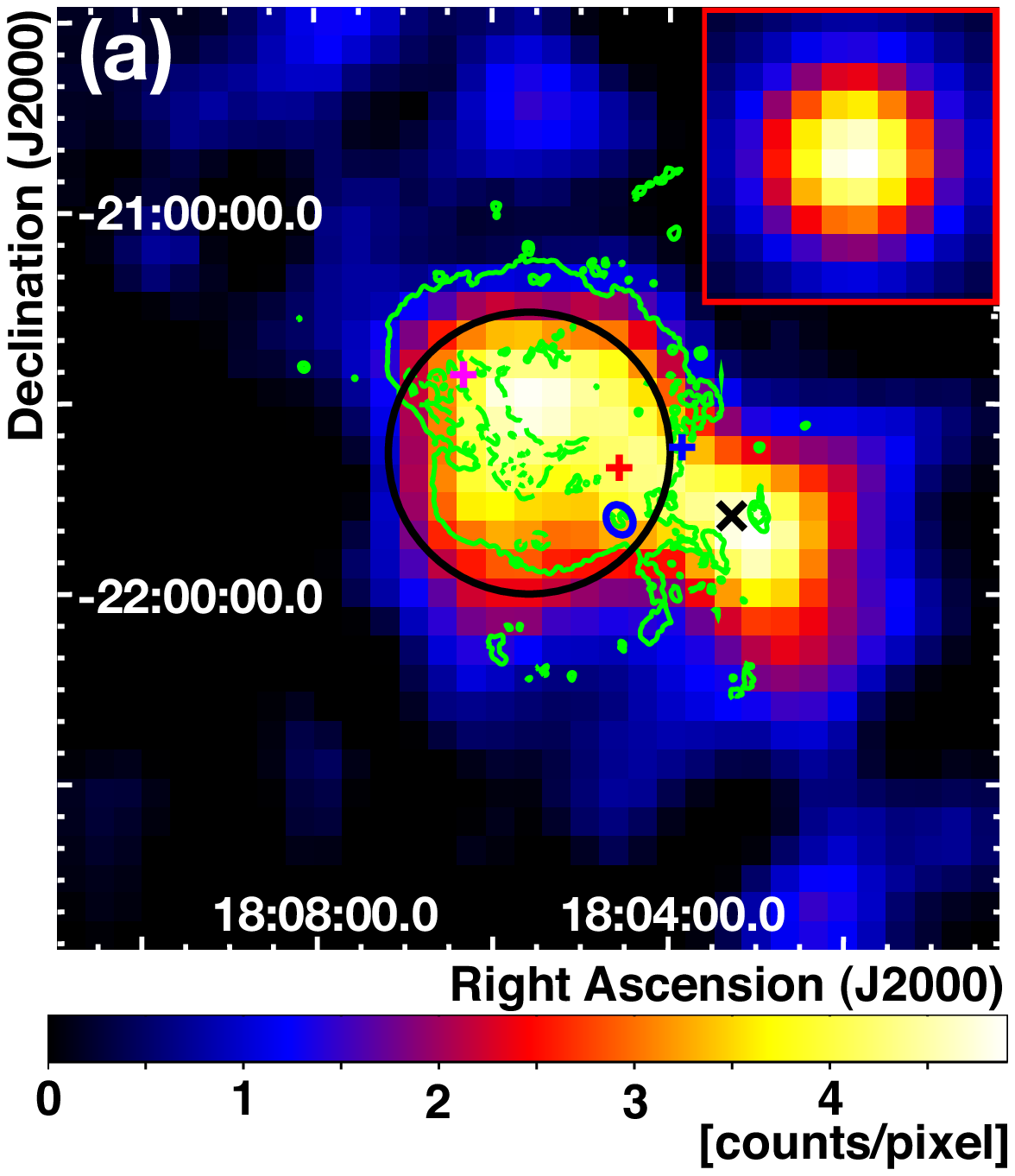}
 \includegraphics[scale=.45]{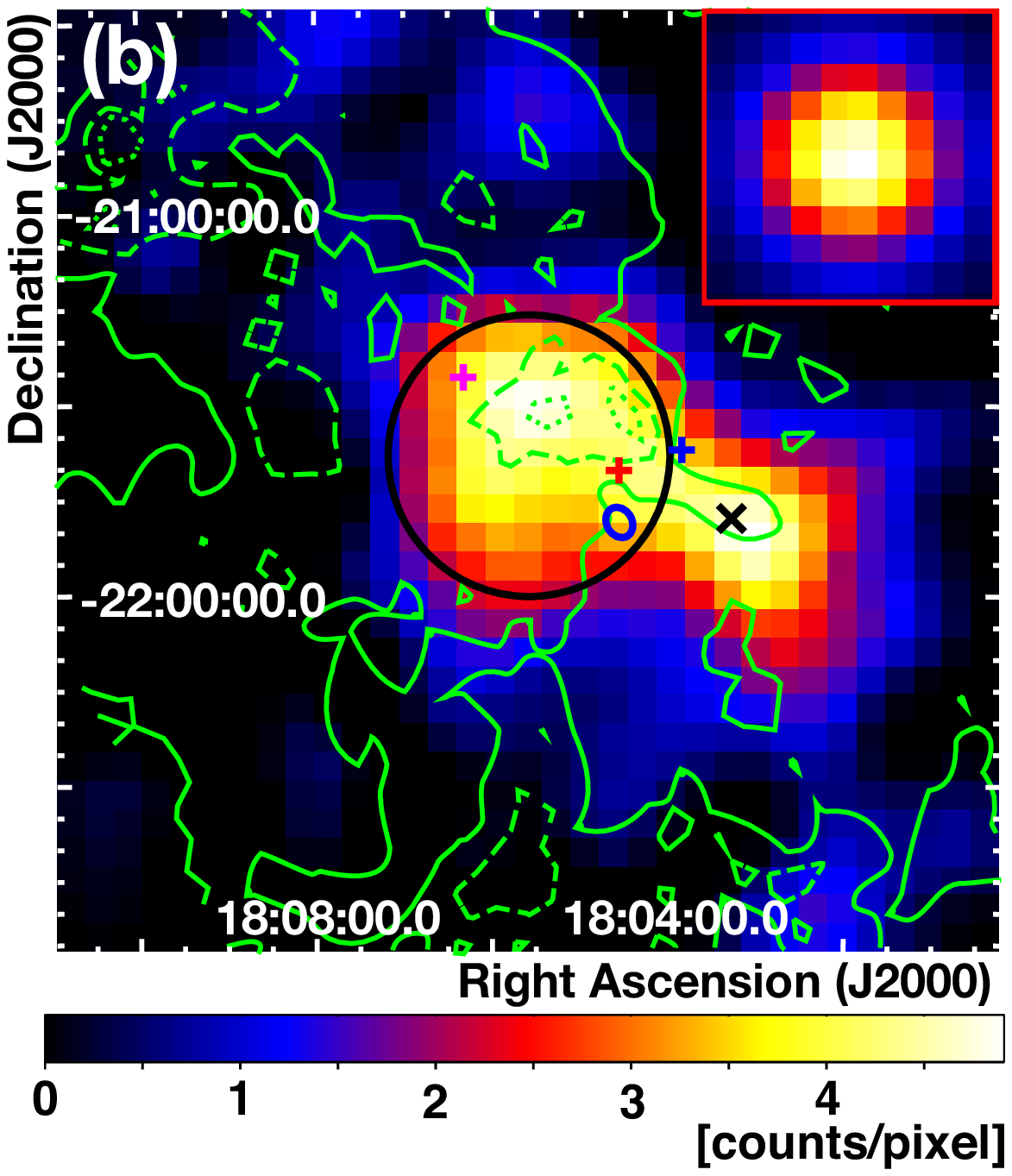}
 \includegraphics[scale=.45]{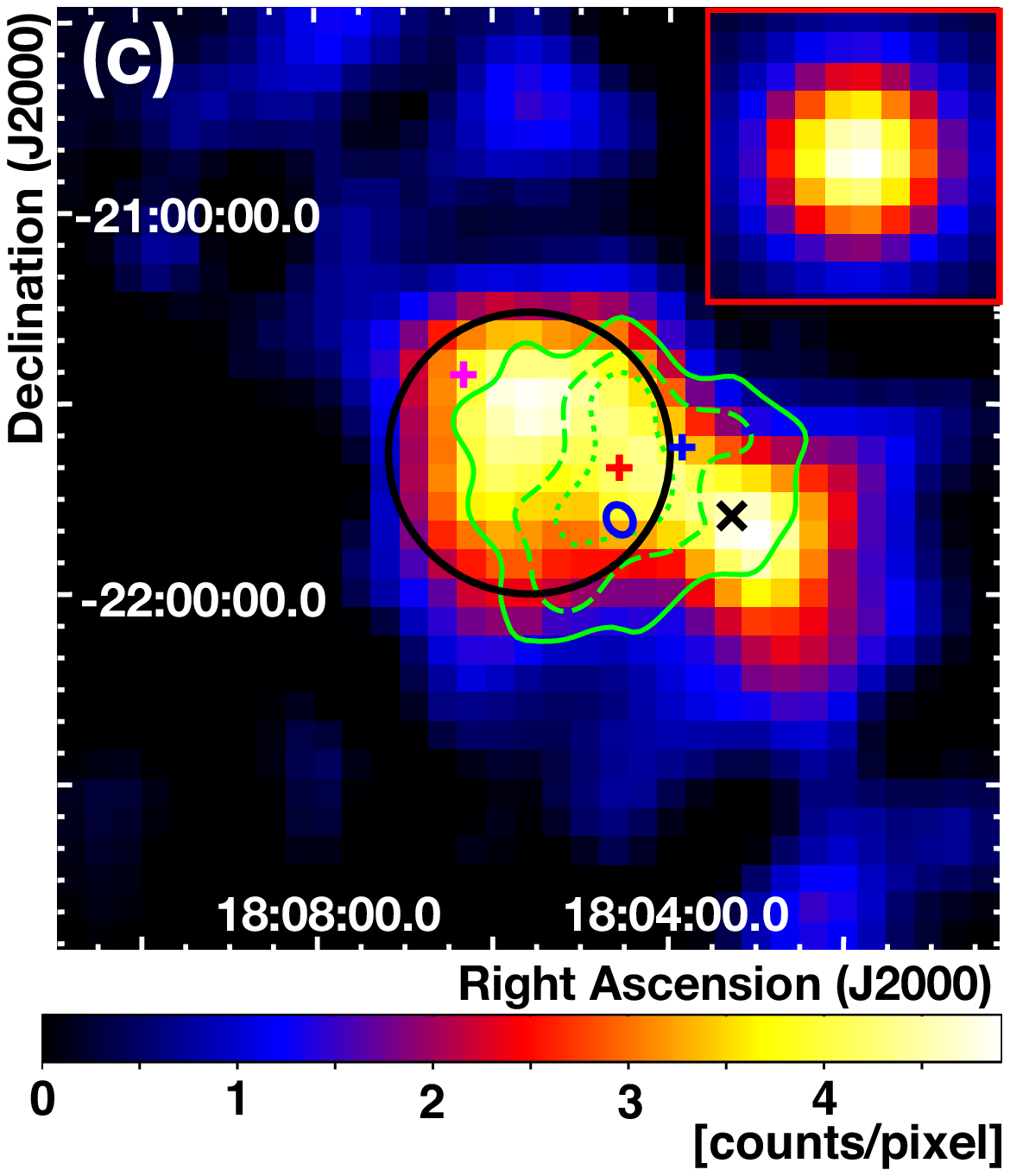}
\caption{Close up views of the LAT 2-10~GeV counts map around G8.7-0.1
 subtracting a fitted diffuse emission model including the isotropic
 component. The counts map has a pixel size of
 0$\fdg$075 and is smoothed by a Gaussian kernel of $\sigma$ $=$ 0$\fdg$225.
The inset of each figure shows the effective LAT PSF for a photon
 spectral index of 2.5. A black circle in the east of each figure
 indicates the best-fit disk size for Source E. A black cross
 indicates the position of Source W. 
A blue and a magenta plus
 indicate PSR J1803$-$2137 and PSR J1806-2125, respectively.
A red plus indicates the PWN candidate Suzaku J1804-2140.
A blue ellipse indicates the radio extension of SNR
 G8.31$-$0.09~\citep{Brogan06}.
Green contours in (a) show the VLA 90 cm image~\citep{Brogan06} at 5,
 15, and 25\% of the peak intensity.
Green contours in (b) give CO~(J~=~1--0) line intensity taken by
 NANTEN~\citep{NANTEN1, NANTEN2} at
 25, 50, 75\% levels, for the velocity range from 20 to 30 km
 s$^{-1}$, corresponding to kinematic distances of approximately 3.5 to 4.5
 kpc. Green contours in (c) indicate the subtracted TeV photon counts
 of HESS J1804$-$216 at 25, 50 and 75\% levels~\citep{HESSsurvey06}.
\label{fig:cmap}}
\end{figure}

Before investigating the extension in detail, we first determined
the strength of the diffuse gamma-ray emission around G8.7$-$0.1 by using the tools
described above. The morphology analysis included only events above 2 GeV to
take advantage of the narrower PSF at higher energies.
We determined the flux and spectral parameters for all model components.
In this process, the normalization of the Galactic
diffuse emission and the flux and spectral index of a power-law model
for the sources within 4$^\circ$ of the direction of G8.7$-$0.1 were set
free to account for the effects of the sources nearest to G8.7$-$0.1 on the
fit. The spectral parameters for more distant sources were fixed to the
values in the 1FGL catalog.
The flux and spectral parameters were then fixed, with the exception of those for the three sources overlapping G8.7$-$0.1.
Figure~\ref{fig:cmap} shows a close-up view of the counts map around G8.7$-$0.1 in the
2--10 GeV band with the diffuse emission subtracted.
Contours representing the radio emission, CO line intensity
and TeV gamma-rays are overlaid on the GeV gamma-ray map.

The coincidence of 1FGL~1805.2$-$2137c and 1FGL~1806.8$-$2109c with the eastern enhancement in the GeV counts map suggested extension of the emission.
To evaluate the presence and size of an extended source, we modeled it
as a uniform disk.
We varied the radial size and centroid of the disk while holding the position
of neighboring 1FGL 1803.1$-$2147c fixed at the catalog values, and evaluated the
resulting maximum likelihood value (\emph{L}$_{\rm ex}$) with respect to
the maximum likelihood for the no-source hypothesis (\emph{L}$_{0}$).
 The largest likelihood ratio,
$-2\ln({L_{0}/L_{\rm ex}})$ (9 degrees of freedom) of $\approx$~478 was
obtained for a disk radius $\sigma$ of 0$\fdg$37.
This is substantially better than that obtained for a model containing two
point sources, $-2\ln({L_{0}/L_{\rm 2s})}$ (12 degrees of freedom)~$\approx$~433,
where $L_{2s}$ is the likelihood for two sources instead of a disk
shape, whose positions were free in the optimization. Therefore, we
conclude that the eastern part of the GeV emission is
significantly extended.
Hereafter, we refer to the emission as
Source~E and employ a uniform disk as the
spatial model for further analysis.
The best-fit centroid for the disk model in J2000 is found to be
(R.A., decl.)~$=$~(18${}^{\rm h}$05${}^{\rm m}$.6,
$-$21${}^\circ$38$'$.0)
with an error radius of 0$\fdg$028 at the 68~\% confidence level.

The extension of the third source, 1FGL 1803.1$-$2147c, (hereafter,
Source~W) was also investigated using the same procedure as above.
We did not find significant extension in that case.
An upper limit on the spatial extent of the gamma-ray emission was obtained by
investigating the decrease of maximum likelihood with increasing radial size of the source in the input emission model. 
Under the assumption of a uniform disk, the upper
limit on the radius was 22$'$ at the 68~\% confidence level.
The best-fit location for Source~W in J2000 was estimated to be
(R.A., decl.)~$=$~(18${}^{\rm h}$03${}^{\rm m}$.3,
$-$21${}^\circ$47$'$.8) with an error radius of 0$\fdg$038 at the
68~\% confidence level.

To quantitatively evaluate the correlation of the GeV emission with other
wavebands, we also performed the likelihood analysis using 
spatial templates derived in those bands in place of the best-fit models derived above.
We calculated the maximum likelihood for
a VLA radio image at 90 cm~\citep{Brogan06} with a point source
added to model Source~W since it does not appear to have a radio counterpart.
We additionally calculated the maximum likelihood using the background-subtracted counts map of TeV gamma rays measured by HESS~\citep{HESSsurvey06}.
To allow for background fluctuations in the VLA and HESS templates,
the fits were performed by changing the extracted regions of the templates
(see Table~\ref{tab:likelihood}).
A simple power-law function was assumed for the spectral models of the above
spatial templates.
Note that we did not use the CO images to form spatial templates since they inevitably contain
large amounts of matter unrelated to the gamma-ray emission from the remnant. 
The resulting maximum likelihood values with respect to the null hypothesis (no emission associated with G8.7$-$0.1) are summarized in Table~\ref{tab:likelihood}.
The likelihood ratio for the radio image is higher than for the model containing three point
sources, while the likelihood ratio for the HESS image is significantly lower. 
Therefore, we conclude that the radio morphology correlates
reasonably well with the GeV emission while the TeV morphology does not.

\subsection{Energy Spectrum}
We used the maximum likelihood fit tool, {\tt gtlike}, for the spectral
analysis of the LAT sources.
In order to produce a spectral energy distribution~(SED) in a model-independent 
manner, fits were performed in eight logarithmically spaced-energy bins covering energies from
200~MeV to 100~GeV.
Within each energy bin we fixed the spectral index at 2 for the LAT sources. Note that the 
flux within an energy bin can vary up to $\sim$~10~\% depending on the choice of spectral index and is taken into consideration as a systematic error.
 
\begin{figure}
\epsscale{.80}
\plotone{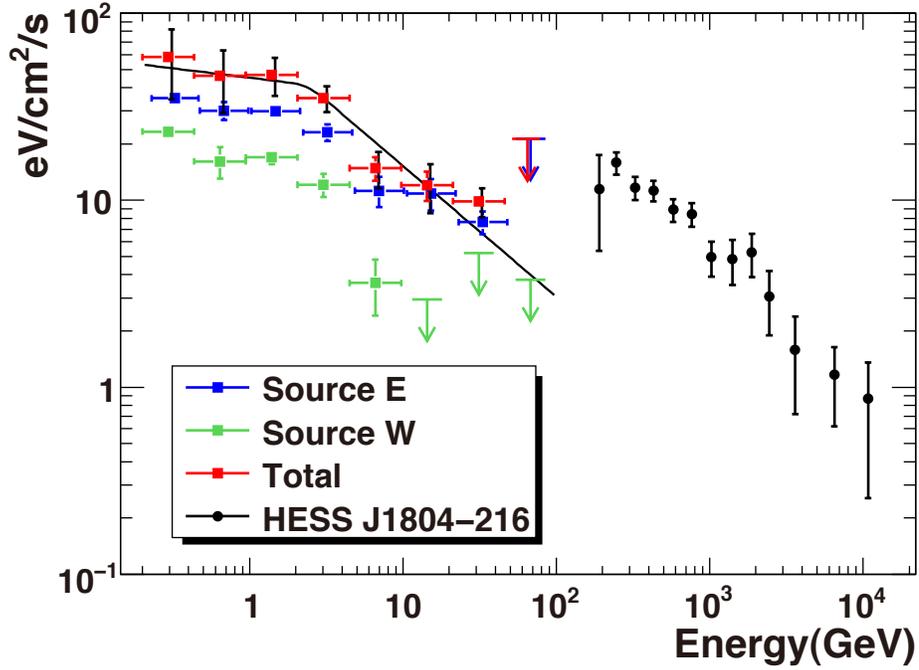}
\caption{Spectral energy distribution of the \emph{Fermi} LAT sources associated with
 G8.7$-$0.1. The blue and green squares with statistical error bars are
 the LAT fits for Source~E and Source~W, respectively.
The red squares in the GeV regime are the total flux of the LAT
 data for both sources. 
Vertical bars in red and in black in the GeV band show the statistical
 errors and systematic errors of the total flux, respectively.
Upper limits are obtained at the 90\% confidence level
 in energy bins in which the likelihood test statistic is $<$ 9 or the number of
 photons predicted by the best-fit model is less than 10. 
The black line shows the best-fit broken power-law model for the total
 spectrum.
The black circles represent data points for
 HESS~J1804$-$216~\citep{HESSsurvey06}.\label{fig:sed}}
\end{figure}

The resulting SEDs for Source~E and Source~W are shown in Figure~\ref{fig:sed}. 
The overlap with the spatially-connected molecular clouds suggests that
there might be a physical connection; thus, we also obtained the total
SED of the two sources together, as shown in Figure~\ref{fig:sed}.

We accounted for systematic errors caused by uncertainties in the
extension, the Galactic diffuse model, and the LAT effective area.
Systematic errors associated with the extension were estimated by varying
the size of Source~E by $\pm$~1$\sigma$.
We considered the energy and positional dependence for the
systematic errors of the Galactic diffuse model. The energy dependence
was estimated by using the residual gamma-ray data with respect to the
best-fit model in a region where no LAT source is present.
We used the neighboring regions on both sides of G8.7$-$0.1 along
the Galactic plane (see Figure~\ref{fig:cmap_wide});
(i) $l$~=~7$\fdg$1-7$\fdg$7 and $b$~=~$-$1$\fdg$1-0$\fdg$9, 
(ii) $l$~=~9$\fdg$15-9$\fdg$75 and $b$~=~$-$1$\fdg$1-0$\fdg$9.
The observed residual can be modeled as
$\sim$~(100~($E$/1~GeV)$^{1.89 \times 10^{-2}}-100$)~\% 
and$\sim$~(101~($E$/1~GeV)$^{-0.89 \times 10^{-2}}-100$)~\% of the total
Galactic diffuse flux for (i) and (ii), respectively.
The normalization of the Galactic diffuse model was adjusted according to the
above equations to estimate the systematic error on the source flux. 
On the other hand, the positional dependence of this residual was
estimated by \cite{W49B} and found to be $\sim$~6\%. We
evaluated the systematic errors due to positional dependence by varying
the normalization of the Galactic diffuse model by $\pm$~6\% from the
best fit value.
We also evaluated systematic errors due to uncertainties in the LAT effective area,
which are 10\% at 100~MeV, decreasing to 5\% at
500~MeV, and increasing to 20\% at 10~GeV and above~\citep{Rando2009}.
The combined systematic errors on the flux are shown by the
black bars in Figure~\ref{fig:sed}.

We evaluated the possibility of a spectral break for the combined
LAT source, i.e., the sum of Source~E and Source~W, in the LAT energy band
by comparing the likelihood of a simple power-law model and a
smoothly broken power-law model for both sources.
The smoothly broken power-law model was described as
\begin{eqnarray}
\frac{dN}{dE}=KE^{-\Gamma_1}\left(1+\left(\frac{E}{E_{\rm break}} \right)^{\frac{\Gamma_2-\Gamma_1}{\beta}} \right)^{-\beta}, 
\label{eq:cutoff}
\end{eqnarray}
where the photon indices, $\Gamma_1$ below the break and $\Gamma_2$ above the
break, the break energy ${E_{\rm break}}$, and the normalization factor
$K$ were free parameters. The parameter $\beta$ was held fixed at 0.05. 
To treat Source~E and Source~W as a combined source, the spectral parameters were varied jointly with the exception of the flux normalizations, which were allowed to vary independently.
The fit yields a likelihood ratio
$-$2~ln($L_{\rm PL}$/$L_{\rm BPL}$)~$\approx 32$, where $L_{\rm PL}$
and $L_{\rm BPL}$ are the likelihoods for the simple power-law model
and the smoothly broken power-law model, respectively. In a worst-case scenario enforcing 1~$\sigma$ systematic uncertainties, the likelihood ratio
decreases to $\sim$ 23 (corresponding to 4.4~$\sigma$
with 2 degrees of freedom).
The best-fit parameters obtained for the smoothly broken power-law model were photon
indices $\Gamma_1 = 2.10 \pm 0.06$ (stat) $\pm$ 0.10 (sys),
$\Gamma_2 = 2.70 \pm 0.12$ (stat) $\pm$ 0.14 (sys),
and $E_{\rm break} = 2.4 \pm 0.6$ (stat) $\pm$ 1.2 (sys) GeV.

We also investigated the spectral shape of each source separately.
The best-fit spectral parameters for Source~E were found to be $\Gamma_1 = 2.10 \pm 0.11$ (stat)
$\pm$ 0.12 (sys), $\Gamma_2 = 2.47 \pm 0.09$ (stat) $\pm$ 0.13 (sys),
and $E_{\rm break} = 1.8 \pm 0.7$ (stat) $\pm$ 0.7 (sys) GeV, while those
for Source~W were found to be $\Gamma_1 = 2.08 \pm 0.15$ (stat)
$\pm$ 0.25 (sys), $\Gamma_2 = 5.88 \pm 1.67$ (stat) $\pm$ 1.27 (sys),
and $E_{\rm break} = 4.5 \pm 0.3$ (stat) $\pm$ 0.1 (sys) GeV.
Thus, we cannot conclude with the present data that the spectral
shapes differ significantly since the break energy
of Source~E and the $\Gamma_2$ of Source~W have large errors.
The smoothly broken power-law models above give
detection significances for Source~E and Source~W in the 200~MeV--100~GeV
of 28~$\sigma$ and 16~$\sigma$, respectively.

We investigated the spectral connection between the GeV and TeV energy
bands. 
Here, we used a chi-squared test for the spectral fit of the LAT
data and the HESS measurements~\citep{HESSsurvey06}.
A fit assuming a broken power-law model yields
a null-hypothesis probability of less than 1.0 $\times$ 10$^{-18}$,
including the worst-case 1~$\sigma$
systematic uncertainties. 
We conclude that the GeV spectrum does
not connect to the HESS measurements smoothly.

\subsection{Time Variability and Pulsation Search of the LAT Source}

Source~W could not be spatially resolved by our work, so other gamma-ray source candidates must be considered. We checked the time 
variability of Source~W to test the
hypothesis of a background active galactic nuclei.
We divided the data into two-month intervals and fit for the flux of Source~W. 
The flux showed no significant time variability, indicating a steady source of emission for this time scale and making it less likely to be a gamma-ray blazar.

We also checked for previously undetected gamma-ray pulsars in this region.
The ATNF
database~\citep{Manchester05}\footnote{http://www.atnf.csiro.au/research/pulsar/psrcat}
lists two nearby radio pulsars with a spin-down power $\dot{E}$ typical of
the known gamma-ray pulsars: PSR J1803$-$2137~\citep{Clifton86} and PSR
J1806$-$2125~\citep{Hobbs02} as shown in Figure~\ref{fig:cmap}.
PSR J1803$-$2137 has $\dot{E}$ = 2.2$\times$10$^{36}$ erg~s$^{-1}$ and a
nominal distance of 3.88~kpc derived from the Dispersion Measure (DM) using
NE2001~\citep{Cordes02}, while PSR J1806$-$2125 has
$\dot{E}$~=~4.3$\times$10$^{34}$
erg~s$^{-1}$ and an estimated distance of 9.85~kpc. Considering the above values, the
latter is not expected to emit a detectable gamma-ray flux.
Using rotational ephemerides provided by the
Parkes~\citep{Weltevrede10}, Nancay~\citep{Theureau05}, and Jodrell
Bank~\citep{Hobbs04} radio telescopes, we phase-folded the gamma-ray
data, but found no evidence for gamma-ray pulsations.

The observed properties of gamma-ray pulsars~\citep{PulsarCatalog} suggest that PSR J1803$-$2137 could have a flux of 9.1$\times$10$^{-8}$ ph~cm$^{-2}$~s$^{-1}$ above 100~MeV.
To check explicitly for the presence of emission from this source, we performed likelihood fits with an additional point source modeled
at the position of PSR J1803$-$2137. The spectral model was assumed to
be a power-law with an exponential
cutoff and the normalization was set free while the spectral index and
cutoff energy were fixed at 1.5 and 2~GeV, the average values
in the Fermi 1st Pulsar catalog~\citep{PulsarCatalog}. 
The likelihood was not improved with the addition of the pulsar to the model.
In the absence of a detection, we set an upper limit on the flux at
the 90\% confidence level of 1.0$\times$10$^{-8}$ ph~cm$^{-2}$~s$^{-1}$ above 100~MeV and conclude that the pulsar does not significantly
contribute to the GeV emission.

\cite{PulsarCatalog} calculated a 5~$\sigma$ flux sensitivity for pulsations to be detected by a blind search of six months of LAT data.
The general upper limit for the Galactic plane of $\simeq$~2.0 $\times$
10$^{-7}$ photons~cm$^{-2}$~s$^{-1}$ above 100~MeV is similar to the
flux of Source W. Therefore, we cannot completely exclude the
possibility that Source W is a gamma-ray pulsar.
The above results in combination with the observed extension of Source~E imply that the bulk of the gamma ray emission near the remnant does not come from an undetected gamma-ray pulsar.

\section{DISCUSSION}
\label{sec:discuss}
\subsection{Origin of the GeV Emission}
\label{sec:GeV}
\subsubsection{Assumptions for Spectral Modeling}
\label{sec:spec_modeling}
We have analyzed the GeV gamma rays in the vicinity of G8.7$-$0.1 and
found the emission to be significantly extended.
The bulk of the emission (Source~E) is positionally coincident with
the synchrotron radio emission from the SNR G8.7$-$0.1,
while a lesser part (Source~W), located outside the western boundary of
G8.7$-$0.1, has no obvious counterpart within the 95\% confidence region obtained by a point source model.
The GeV morphology is reasonably well represented by the radio emission of
the SNR, suggesting a correlation with high-energy electrons.
There are molecular clouds spatially
associated with G8.7$-$0.1~\citep{Blitz82} and likely to be interacting with 
the SNR since an OH maser is detected on the
eastern edge of G8.7$-$0.1~\citep{Hewitt09}. 
The GeV gamma rays overlap with these spatially-connected molecular
clouds.
This implies a physical connection of the two LAT sources although there
remains a possibility that Source~W is a gamma-ray pulsar undetected
at the current sensitivity.
Here, we assume that the bulk of GeV gamma-rays comes from
the interaction between particles accelerated by the SNR and gas in the
clouds, where the particles are confined in the SNR shell.
Also, we assume that the molecular clouds uniformly cover the whole
surface of the SNR since the CO emission is not significantly localized
in any part of the SNR.
Note that since
the eastern part of this region dominates the GeV emission,
the contribution of the western source does not affect any conclusions
drawn in this paper. We discuss the possibility that the GeV gamma-rays
come from other sources in Section~\ref{sec:OtherSources}.

The TeV gamma-ray source, HESS~J1804$-$216, overlaps
the GeV gamma rays.
However, the TeV morphology is not better correlated with the GeV
gamma rays than the radio emission from G8.7$-$0.1 and the GeV spectrum
does not connect smoothly to the TeV spectrum, indicating another emission component, either additional high-energy particles accelerated by G8.7$-$0.1
or a high-energy source unrelated to G8.7$-$0.1. In
section~\ref{sec:GeV}, we focus on the GeV emission and reserve discussion of the relation
between the TeV and GeV emission for Section~\ref{sec:TeV}.

Below, we adopt the simplest assumption, that a population of
accelerated protons and electrons is distributed in a region
characterized by constant density and magnetic field strength and the
injected electrons have the same momentum distribution as the protons.
This assumption implies a break in the particle momentum spectrum
because the spectral index of the radio data, which corresponds to lower
particle momenta, is much harder than that of the gamma-rays, which
correspond to higher particle momenta.
We use the following functional form to model the momentum
distribution of injected particles:
\begin{eqnarray}
Q_{e,p}(p)=a_{e,p}\left(\frac{p}{1~\mathrm{GeV}~c^{-1}}\right)^{-s_{\rm L}}\left(1+\left(\frac{p}{p_{\rm br}}\right)^2\right)^{-(s_{\rm H}-s_{\rm L})/2},  
\end{eqnarray}
where $p_{\rm br}$ is the break momentum, $s_{\rm L}$ is the spectral
index below the break and $s_{\rm H}$ above the break. 
Note that here we consider the minimum momenta of protons and electrons
to be 100~MeV~$c^{-1}$ since the details of the proton/electron
injection process are poorly known.

Electrons suffer energy losses due to ionization (or Coulomb
scattering), bremsstrahlung, synchrotron processes, and IC scattering.
The modification of the electron spectral distribution due to such
losses was calculated according to~\cite{Atoyan95}, where electrons are
assumed to be injected at $t = 0$ from an impulsive source.
Since diffusive shock acceleration theory generally predicts particle
acceleration in the Sedov phase with a typical duration
$10^3$--$10^4$~yr, the assumption of an impulsive source is a
good approximation for G8.7$-$0.1, which has an age of $2.5 \times 10^4$~yr.
Note that we ignore the radiative cooling of protons since the time
scale of the energy loss due to nuclear interaction is $\approx$ 6
$\times$ 10$^{7}$ (1 cm$^{-3}$/$\nh$) yr \citep{Aharonian96}, which is
much greater than the SNR age unless the environment is very dense.
We adopt 4~kpc for the distance from the Earth to the SNR since the
GeV emission overlaps molecular clouds corresponding to
a kinematic distance of 3.5--4.5~kpc.
The gamma-ray spectrum from $\pi^0$ decays produced by the interaction of
protons with ambient hydrogen is scaled by a factor of 1.84 to account
for helium and heavy nuclei in the target material and the cosmic-ray
composition~\citep{Mori09}. 
We also consider the contribution of the emission from secondary
e$^{+}$/e$^{-}$ produced by charged pion production and decay in
the $\pi^{0}$ decay model.
For the calculation of spectra of the secondaries, we use the parametric
formulae from~\cite{Kamae06}.

\subsubsection{SNR G8.7$-$0.1}
\label{sec:SNR_Origin}
First, we consider a $\pi^0$-decay model to account for the broadband
gamma-ray spectrum.
We use an electron-to-proton ratio of $K_{ep} = 0.01$, which is the
ratio found in the local cosmic-ray abundance. Here the ratio is defined
at a particle momentum of 1~GeV~$c^{-1}$.
The spectral index of proton momentum in the high-energy regime is constrained to
be $s_{\rm H} \approx 2.7$ from the gamma-ray spectral slope. 
The index below the break is determined to be $s_{\rm L}\approx
2.0$ by the observed radio spectrum due to synchrotron radiation by
relativistic electrons~\citep{Kassim90}.
Note that we do not account for the spectral turnover at radio frequencies
which is due to absorption by localized
thermal gas associated with one or more of the \ion{H}{2} regions near the W30
complex~\citep{Kassim90}.
The observed gamma-ray luminosity requires the gas density to be much
larger than $\nh\approx 1\ \mathrm{cm}^{-3}$ averaged over the entire
SNR shell in order not to exceed the typical kinetic energy of a supernova
explosion~($\sim 10^{51}\ {\rm erg}$).
We assume $\nh = 100\ \mathrm{cm}^{-3}$ which is a
typical value in molecular clouds.
This is consistent with CO observations because the mass of a shell
of radius 26~pc\footnote{This is estimated with the apparent diameter of the SNR and the distance of G8.7$-$0.1 to the Earth.}
and thickness 7~pc\footnote{This is constrained by the apparent
distance from the edge of the radio emission to the position of Source~W.}
with density $\nh$ is 1.9 $\times$ 10$^{5}$ M$_{\odot}$ which is
smaller than the value of 5.9 $\times$ 10$^{5}$ M$_{\odot}$ estimated
from the CO data taken by the NANTEN telescope~\citep{NANTEN1,
NANTEN2}.

\begin{figure}
\epsscale{0.8}
\plotone{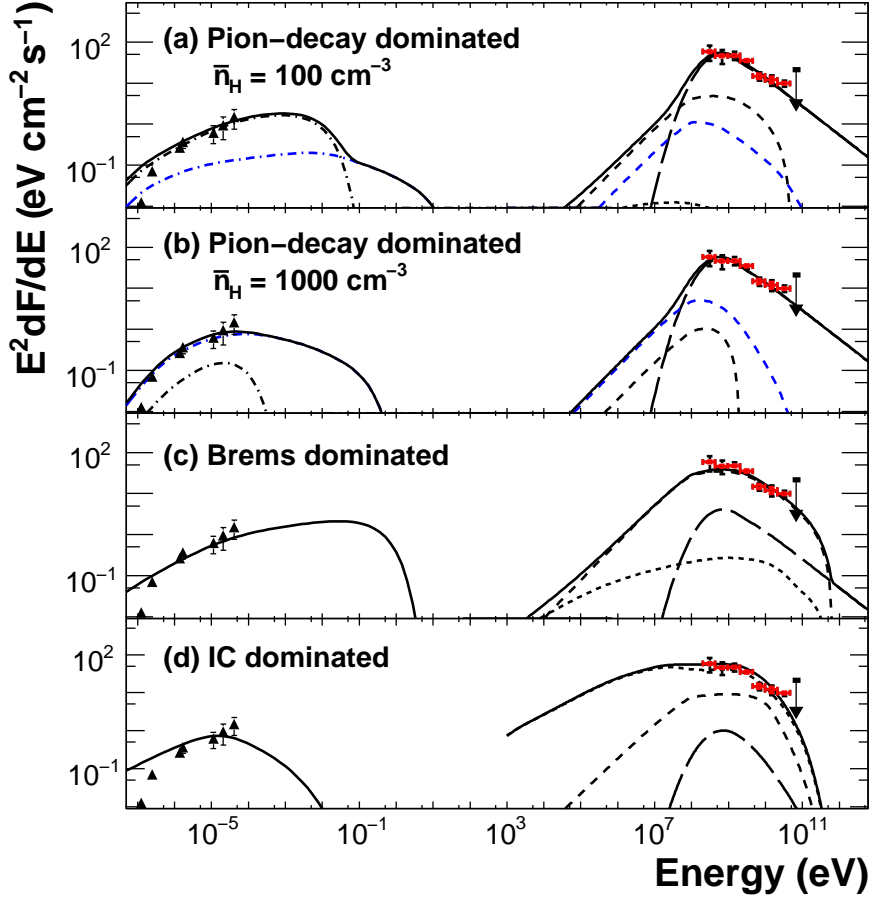}
\caption{
Multi-band spectra of the \emph {Fermi} LAT emission associated with SNR
 G8.7$-$0.1. The red squares in GeV regime are the LAT data, where
the red and black bars are the statistical and systematic
 errors, respectively. 
The radio emission from the entire region of G8.7$-$0.1~\citep{Kassim90} is modeled by synchrotron
 radiation, 
while the gamma-ray emission is modeled by different combinations of $\pi^0$-decay~(long-dashed curve),
bremsstrahlung~(dashed curve), and inverse Compton~(IC)
 scattering~(dotted curve). 
In the panel of (a), (b) and (c), the blue curves show the emission from the
 secondary e$^{+}$/e$^{-}$, and the dot-dashed
 lines in the radio  band show the
 emission from the primary electrons and secondary e$^{+}$/e$^{-}$.
Details of the models are described in the text.
 \label{fig:spec}}
\end{figure}

Using the parameters summarized in Table~\ref{tab:model},
we calculated radiation model curves as shown in Figure~\ref{fig:spec}~(a).
The resulting total proton energy, $W_{p}\sim 2.8 \times 10^{49}\cdot(10^2\
\mathrm{cm}^{-3}/\nh)\cdot(d/4\ \mathrm{kpc})^{2}$~erg, is less than
10\% of the typical kinetic energy of a supernova explosion and quite
reasonable. 
Note that $W_{p}$ is not the total energy of accelerated
protons but that of the in--situ protons in the molecular clouds.
$W_{p}$ is changed up to a factor of $\sim$ 2 within the
uncertainty of its distance (3.2--6~kpc), which does not affect our
conclusion.
It is difficult to derive the break point of the proton momentum spectrum from the break in the gamma-ray spectrum since it lies in the region where we expect a gamma-ray spectral curvature due to the kinematics of $\pi^0$ production and decay.
The gamma-ray spectrum gives an upper bound for the
momentum break at $\sim$~10~GeV~$c^{-1}$.
The momentum break cannot be lower than $\sim$~3~GeV~$c^{-1}$ to avoid
conflict with the radio data.
Here we adopt 3~GeV~$c^{-1}$.
The magnetic field strength is constrained to be $B \sim 100\;\mu$G,
which is plausible since a magnetic field can be amplified up to several
hundred $\mu$G by the
compression of gas by the shock of a middle-aged SNR in molecular
clouds~\citep{Chevalier1999}.
The emission from secondary e$^{+}$/e$^{-}$ does not significantly
contribute to the emission from G8.7$-$0.1 in this modeling given the
above $\nh$.

To consider the
situation when the synchrotron emission from the secondaries dominates in
the radio band, we model the GeV emission with $\nh = 1000\ \mathrm{cm}^{-3}$.
Although the mass of the molecular clouds with the assumed shape
exceeds the value estimated by using the NANTEN data in this case, that $\nh$ 
is possible if the clouds responsible for the GeV emission have
nonuniform structure. The modeling results are shown in
Table~\ref{tab:model} and Figure~\ref{fig:spec}(b), respectively.
The range of the momentum break changes slightly to
3--15~GeV~c$^{-1}$ and the magnetic field is constrained to be
$\sim$ 400~$\mu$G. From the above considerations, the $\pi^{0}$ decay
model can explain the GeV emission although the magnetic field and the
momentum break depend on $\nh$.

Leptonic models struggle to match the GeV gamma-ray spectrum.
The electron-bremsstrahlung dominated model (Figure~\ref{fig:spec}(c))
cannot explain the GeV gamma rays unless $K_{ep}$ of the accelerated particles is 
much larger than $\sim$ 0.01, as found in the local cosmic-ray abundance.
For the inverse Compton (IC) dominated model (Figure~\ref{fig:spec}(d)),
the total energy in electrons is calculated to be
$W_{e}\approx9.9 \times 10^{50}\cdot(d/4\ \mathrm{kpc})^{2}$~erg, 
for an energy density of $\sim$~2.9~eV~cm$^{-3}$ for the interstellar 
radiation field\footnote{The interstellar radiation field (see Table~\ref{tab:model})
for non cosmic microwave background at the location of G8.7$-$0.1 is estimated from the GALPROP code~\citep{Porter08} with approximation of two infrared and two optical blackbody components.}, 
which is comparable to the typical kinetic energy of a supernova explosion ($\sim$ 10$^{51}$ erg).
Therefore, the IC dominated model is not plausible
unless the radiation field is at least 10 times more intense than
expected.

Assuming the $\pi^{0}$-decay model, our observations of the LAT source in
the vicinity of G8.7$-$0.1 when combined with the radio data constrain the
proton momentum break to be in the range 3--15~GeV~c$^{-1}$.
This spectral feature might indicate the escape of the accelerated
particles confined around the blast waves propagating into the dense
clouds \citep[e.g.,][]{Uchiyama10}.
On the other hand, \cite{Gabici07} discussed the time evolution of
non-thermal emission from molecular clouds illuminated by cosmic rays
from a nearby SNR and predicted a steep gamma-ray spectrum for an old
SNR due to energy-dependent diffusion of cosmic rays.
The two models do not produce the emission at the same place. In the
former, the gamma-ray emission comes from the cloud shock, which can 
be traced by non-thermal radio filaments. However,
the spatial difference is too small to be resolved by the
LAT. Therefore, we cannot say which model is favorable for the GeV emission.

\subsubsection{Other Sources}
\label{sec:OtherSources}
First, we consider the possibility that the GeV gamma rays come from
pulsar wind nebulae (PWNe).
The IC emission from relativistic electrons is the most
plausible process for the gamma-ray emission from
PWNe~\citep{Crab,Vela-X,MSH,Slane10,HESSJ1825}.
A PWN nearer to the Earth than the location of G8.7$-$0.1 is a
possible candidate since the smaller distance loosens the constraint on
energetics derived from an IC dominated model.
There are several known or suspected PWNe in this field, one associated with PSR
J1803$-$2137~\citep{Kargaltsev07} and another a PWN candidate found with Suzaku,
J1804$-$2140~\citep{Bamba07}, which were found only in the X-ray band.
If the IC emission contributes significantly to such a bright GeV
source, then the synchrotron emission in the radio band due to the
corresponding electrons would be detectable unless the magnetic field is weaker than $\sim$~1 $\mu$G.
Magnetic fields in PWNe with GeV emission are estimated to be at
least $\sim$ 3~$\mu$G using the {\emph Fermi} LAT
observations~\citep{Crab,Vela-X,MSH,Slane10,HESSJ1825}.
Thus, we suppose that the bulk of the GeV emission does not come from
a PWN.

Next, we consider the contribution of SNR G8.31$-$0.09,
which has a small size of 5$' \times $4$'$~\citep{Brogan06}.
This SNR is not located at the bright portion of the GeV emission.
Therefore, we conclude that gamma-ray emission from G8.31$-$0.09 cannot
significantly contribute to the GeV emission.
From these considerations, the bulk of the GeV gamma rays is most
naturally explained by the decay of $\pi^{0}$s produced by the interaction of
G8.7$-$0.1 with molecular clouds.

\subsection{Origin of the TeV Emission}
\label{sec:TeV}
\subsubsection{Pulsar Wind Nebula}
\label{sec:TeV_PWN}

We consider the relation between the TeV gamma-ray source,
HESS~J1804$-$216, and the GeV emission.
One possibility is that the TeV emission
arises via the IC scattering of the relativistic electrons in a PWN.
The sizes of the PWNe found in the X-ray, which have extension of at
most $\sim$2$'$~\citep{Kargaltsev07,Bamba07}, are $\sim$ 40
times smaller than the extension of the TeV gamma-ray source.
TeV emission from the IC process can be more extended than
synchrotron X-ray emission from a PWN due to differences in radiative cooling times for the electrons generating the
emission in those bands~\citep{Jager08}.
In the case of the PWN associated
with PSR J1803$-$2137, the extended TeV gamma-ray emission
($\sim$ 30~pc\footnote{The extension is estimated by the apparent size
and the distance of 3.88~kpc from the Earth to the
pulsar}) can be explained if the transporting velocity for
the TeV-emitting electrons 
averaged over the age of the pulsar (15.6~kr;~\cite{Brisken06}) is larger 
than $\sim$ 1900 km s$^{-1}$.
Diffusion parallel to a magnetic field, or convection might explain such
a large propagation velocity~\citep{Jager08}.
On the other hand, the Suzaku J1804$-$2140 PWN is not well-studied and its
origin remains unclear. 
Therefore, we cannot rule out the possibility of a PWN origin for the TeV emission.

\subsubsection{Cosmic Rays Accelerated in G8.7$-$0.1}

Another possibility is that the TeV emission also originates in
the particles accelerated in G8.7$-$0.1. 
It is predicted that TeV gamma-ray emission can arise from the interaction of
cosmic rays that have escaped from an SNR with nearby molecular clouds, say
within $\sim$~100~pc~\citep[e.g,][]{Aharonian96}.  
On the other hand, we argue in Section~\ref{sec:GeV} that the GeV emission comes from the
interaction of particles confined in the shell
of G8.7$-$0.1 with molecular clouds.
A combined scenario can explain the concave spectral shape in the GeV--TeV
band, i.e., the molecular clouds distant from G8.7$-$0.1 along the
line of sight emit the TeV gamma rays, while the molecular clouds
responsible for the GeV emission are located adjacent to the SNR.
The spectral index of 2.72~$\pm$~0.06 for the TeV gamma
rays~\citep{HESSsurvey06} is consistent with the particle spectral index
predicted by a theory assuming the energy-dependent diffusion of
particles accelerated in an SNR~\citep[e.g.,][]{Aharonian96}, supporting
the above scenario.
If this is correct, we can constrain the diffusion coefficient.

We performed the modeling for the GeV and TeV gamma-ray
spectra considering the above scenario. 
Again, we treat G8.7$-$0.1 as an impulsive source injecting
the accelerated particles at $t = 0$.
In addition, we assume that the accelerated particles do not escape into
interstellar space until the SNR enters the Sedov
phase at $t_{Sedov}$.
Under the above assumptions, the density spectrum of diffused protons
is derived by \cite{Gabici09} as: 
\begin{eqnarray}
\label{eq:diffusion}
f(E,R,t)~=~\frac{N_{0}E^{-s}}{\pi^{3/2}R^{3}_{\rm
 diff}}\exp\left(-\frac{R^{2}}{R^{2}_{\rm diff}}\right)~{\rm GeV}^{-1}~{\rm
 cm}^{-3},
\end{eqnarray}
where R is the distance from the center of the SNR and the injection
spectrum is assumed to be a power-law, Q~$\propto$~$E^{-s}\delta{\rm
(R)}\delta{\rm(t)}$. We also adopt Eq. (\ref{eq:diffusion}) for electrons.
The energy losses for protons and electrons are considered as
described in Section \ref{sec:spec_modeling}.
N$_{0}$ is the normalization and proportional to the total proton
energy W$_{tot}$ injected at $t~=~0$ from the source.
R$_{\rm diff}$ is the diffusion radius represented by
2$\sqrt{D({\rm E})(t-{\chi}(E))}$. D($E$) is the diffusion coefficient
described by the following equation,
\begin{eqnarray}
\label{eq:diff_coe}
D(E)~=~D_{10}(E/10~{\rm GeV})^{\delta}~{\rm cm^{2}~s^{-1}},
\end{eqnarray}
where D$_{10}$ is the value of the diffusion coefficient at $E$~=~10~GeV.
$\chi$($E$) represents the confinement of the particles until
t$_{Sedov}$,
where $\chi$($E$) = t$_{Sedov}$~($E/E_{\rm
max}$)$^{-1/\epsilon}$. 
$\epsilon$ determines the release time of the particles with energy $E$.
In the case of $t-\chi(E)$ $\leq$ 0, Eq. (\ref{eq:diffusion}) becomes 0,
i.e., particles are not released from the shell.
We assume t$_{Sedov}$ = 200 yr and $E_{\rm max}$ =
5 PeV, where $E_{max}$ is the highest energy of the accelerated
particles in the SNR. These values vary depending on the
environment of an SNR. However, this does not significantly affect our
results.

For the formalism of Eq. (\ref{eq:diffusion}), the injected
particles below the threshold energy defined by $t-\chi(E) = 0$
remain completely within the SNR shell, while the rest are entirely
released. This is not consistent with the particle spectrum obtained
by the modeling of the GeV spectrum as described before; 
therefore, we extrapolate the
particle spectrum above the threshold energy by using the
broken power-law model. The index above the threshold is the same as the
particle momentum spectrum used in Figure~\ref{fig:spec}(a).
This requires that the momentum break energy ($\sim$ 3 GeV
c$^{-1}$) is consistent with the energy threshold of the escaped 
particles at the SNR age.
Thus, $\epsilon$ is obtained to be 3.0. 
This approximation for the GeV spectrum reduces the
amount of escaped particles just above the
threshold compared to Eq.~(\ref{eq:diffusion}), but it does not
greatly affect the modeling of the TeV emission since the energies of
the contributing particles are much higher than the threshold.

\begin{figure}
\epsscale{0.8}
\plotone{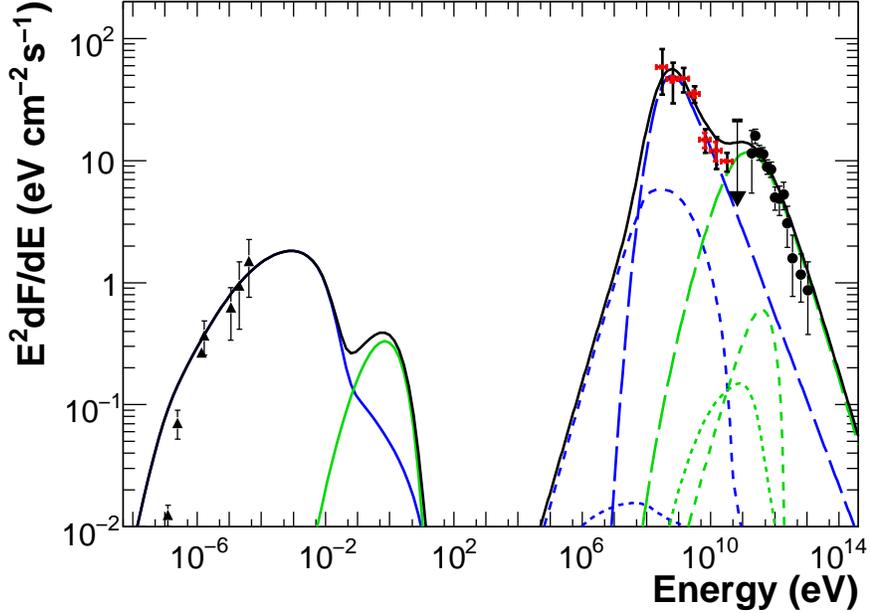}
\caption{
Comparison of the model results for the GeV and TeV gamma-rays with
 the observed spectra.
The red squares in GeV regime are the LAT data, where the red and black
 bars are statistical and systematic errors, respectively.
The black triangles show the radio emission from the entire region of
 G8.7$-$0.1~\citep{Kassim90}.
The TeV emission from HESS~J1804$-$216~\citep{HESSsurvey06} is shown by
the black circles.
The solid black curve shows the total emission of the gamma-rays.
The blue curves show the emission from the molecular clouds responsible for
 the GeV emission, which is a sum of emission from primary e$^{-}$ and secondary  e$^{-}$/e$^{+}$.
The green curves show the emission from the TeV-emitting clouds.
The radio emission is modeled by synchrotron radiation (solid curves), 
while the gamma-ray emission is modeled by different combinations of
 $\pi^0$-decay~(long-dashed curve), bremsstrahlung~(dashed curve), and
 inverse Compton~(IC) scattering~(dotted curve). 
\label{fig:spec_GeV_TeV}}
\end{figure}

We calculate the radiation model curves for the GeV emission with the
same parameters as those of Figure~\ref{fig:spec}(a).
For the TeV emission, the radiation curves are calculated assuming
$K_{ep}$~=~0.01
and the typical values in a molecular cloud for the magnetic
field of 10~$\mu$G and $\nh = 100\ \mathrm{cm}^{-3}$.
The obtained radiation curves are shown in Figure~\ref{fig:spec_GeV_TeV}. 
The amount of secondary e$^{+}$/e$^{-}$ in the TeV-emitting clouds
depends on $\nh$ of gas in which protons, i.e., parent particles of the 
secondaries, propagate, which is uncertain.
If the gas is much denser than $\nh = 100\ \mathrm{cm}^{-3}$,
then the emission from secondary e$^{+}$/e$^{-}$ can
contribute significantly to the radio spectrum. 
However, the resulting parameters from the modeling with the extremely dense gas are not largely affected with the exception of the magnetic field.
Therefore, we neglect the contribution of emission from the
secondaries in the TeV emission.
To simplify the electron energy losses during the propagation,
we use the constant magnetic field and $\nh$ of the TeV-emitting clouds.
This assumption affects only the peak energy of synchrotron emission,
which is not constrained by any observations.

The value of $\delta$ is constrained to be 0.6 by fitting the particle spectrum to
the TeV gamma-ray spectral slopes above the SED peak.
D$_{10}$ is constrained by the cutoff energy of the particle spectrum
corresponding to the SED peak of the TeV emission since the cutoff energy is
proportional to 
R$_{\rm TeV}^{2}$/R$_{\rm diff}^{2}$~=~R$^{2}_{\rm
TeV}$/[4D$_{10}(E/{10~{\rm GeV}})^{\delta}$ ($t-\chi(E)$)],
where R$_{\rm TeV}$ is the distance to the TeV-emitting clouds.
A lower limit on R$_{\rm TeV}$ can be provided
by the radius of the radio shell of $\approx$ 26 (d/4~kpc)~pc
since the TeV-emitting clouds should be located further from the remnant than the GeV-emitting clouds.
As a result, the lower limit on D$_{10}$ is obtained to be
7.5$\times$10$^{25}$~(d/4~kpc)$^{2}$~cm$^{2}$~s$^{-1}$.
The observed differential flux of the TeV emission is
described by 
F$_{\rm TeV}$~$\propto$~W$_{tot}$D$_{10}^{-3/2}10^{3{\delta}/2}$M$_{\rm TeV}$/4${\pi}$d$^{2}$ using
Eq. (\ref{eq:diffusion}) and (\ref{eq:diff_coe}), where M$_{\rm TeV}$ is
the mass of the clouds responsible for the observed TeV emission. Using the above
relation, we can obtain an upper limit on D$_{10}$ from the mass
obtained by the CO(J~=~1$-$0) data with NANTEN. 
We searched for the TeV-emitting clouds in the velocity range from 10 to
40~km~s$^{-1}$, corresponding to the distance to G8.7$-$0.1
and found molecular clouds with the mass of about 2.0~$\times$~10$^{6}$
M$_{\odot}$ for the velocity range from 10 to 30~km~s$^{-1}$. Thus, an
upper limit on D$_{10}$ is 
5.4~$\times$~10$^{26}\cdot$[(W$_{tot}$/10$^{50}$
erg)$\cdot$ (4~kpc/d)$^{2}$]$^{2/3}\cdot$(10$^{({\delta}/0.6)}$/10)
cm$^2$~s$^{-1}$.
In this case, R$_{\rm TeV}$ comes to $\sim$ 70~pc.
The constrained range for D$_{10}$ is much smaller than that obtained by
\cite{Delahaye08}, where $\delta$ and D$_{10}$ were estimated to be
0.46--0.70 and 0.6--6.7~$\times$~10$^{28}$~cm$^{2}$~s$^{-1}$ by using
the observed ratios of secondary to primary nuclei.
However, our results probably represent an environment of dense interstellar gas 
since a lower D$_{10}$~$\sim$~$10^{26}~\mathrm{{cm}^{2}~s^{-1}}$ is expected in that case~\citep{Ormes88}.

We also consider other possible scenarios: 1) both the GeV and
TeV gamma rays originate from the interaction escaped particles accelerated in G8.7$-$0.1 with molecular clouds,
2) the GeV emission arises from the mechanism predicted
by~\cite{Uchiyama10}.
To examine the possibility of scenario 1), we perform the modeling
for the GeV and TeV emission using Eq.~(\ref{eq:diffusion}).
As a result, D$_{10}$ is constrained to be 4.0 $\times$ 10$^{27}$
(d/4 kpc)$^2$ cm$^{2}$~s$^{-1}$ by the cutoff energy of the particle spectrum
corresponding to the SED peak of the GeV emission, while $\delta$ is
constrained to be 0.6 by fitting the particle spectra to the gamma-ray
spectral slopes above the SED peaks. 
The model of the TeV emission using
the obtained diffusion coefficient gives M$_{\rm TeV}$
$=$ 4.4 $\times$ 10$^{7}$ M$_{\odot}$, which is much larger than the observed
value. Therefore, this scenario is unlikely. 
In the case of scenario 2), \cite{Uchiyama10} state that TeV emission
would not be explained by this mechanism and may instead arise from the interaction of particles that escaped from SNR shocks at earlier epochs with the molecular clouds. Thus, D$_{10}$ does not change.

\section{CONCLUSIONS}
\label{sec:conclusion}
 
We have investigated the GeV gamma rays in the vicinity of the SNR
G8.7$-$0.1 and found that they are extended. 
Most of the emission (Source~E) is positionally coincident with
the SNR G8.7$-$0.1, while a lesser portion (Source~W), located outside the
western boundary of G8.7$-$0.1, has no evident counterpart in other
wavelengths within the 95\% confidence region obtained using a point source
model.
The GeV gamma rays coincide with spatially-connected molecular
clouds, implying a physical connection between the two sources.
The decay of $\pi^{0}$s produced by particles accelerated in the SNR and hitting
the molecular clouds naturally explains the GeV gamma-ray
spectrum since a direct interaction between G8.7$-$0.1 and the molecular
clouds is supported by the detection of an OH maser, although electron
bremsstrahlung cannot be ruled out.

On the other hand, the GeV morphology is not well represented by the TeV
emission from HESS~J1804$-$216.
The GeV gamma-ray spectrum has a break around 2~GeV and falls below
the extrapolation of the TeV gamma-ray
spectrum of HESS~J1804$-$216.
The TeV spectral index is most naturally explained by a theory
assuming the energy-dependent diffusion of particles accelerated in the
SNR, although the possibility that the TeV emission might come from
a PWN cannot be ruled out.
Under the assumption that the bulk of the TeV gamma rays comes from the
interaction between distant molecular clouds and cosmic rays released
and diffused from G8.7$-$0.1, we can constrain the diffusion coefficient
of the particles.

\acknowledgments
The \textit{Fermi} LAT Collaboration acknowledges generous ongoing support
from a number of agencies and institutes that have supported both the
development and the operation of the LAT as well as scientific data analysis.
These include the National Aeronautics and Space Administration and the
Department of Energy in the United States, the Commissariat \`a
l'Energie Atomique and the Centre National de la Recherche Scientifique
/ Institut National de Physique Nucl\'eaire et de Physique des
Particules in France, the Agenzia Spaziale Italiana and the Istituto
Nazionale di Fisica Nucleare in Italy, the Ministry of Education,
Culture, Sports, Science and Technology (MEXT), High Energy Accelerator
Research Organization (KEK) and Japan Aerospace Exploration Agency
(JAXA) in Japan, and the K.~A.~Wallenberg Foundation, the Swedish
Research Council and the Swedish National Space Board in Sweden. 

Additional support for science analysis during the operations phase is
gratefully acknowledged from the Istituto Nazionale di Astrofisica in
Italy and the Centre National d'\'Etudes Spatiales in France.

\begin{table}
\begin{center}
\caption{Likelihood ratios for the different spatial models compared
 with the null hypothesis, no gamma-ray emission from
 G8.7$-$0.1 (2--10 GeV).\label{tab:likelihood}}
\begin{tabular}{lccc}
\tableline\tableline
 Model  & $-2\ln({L_{0}/L})$\tablenotemark{a} &  Additional degrees of
 freedom \\
\tableline
Null hypothesis & 0 & 0 \\
3 point sources\tablenotemark{b} & 433.4 & 12 \\
VLA 90 cm\tablenotemark{c} + Source W & 436.5--462.4 & 6 \\
HESS\tablenotemark{c} & 404.8--408.0 & 2 \\
Uniform disk and point source & 477.8 & 9 \\
PSR~J1803$-$2137\tablenotemark{d} & 477.8 & 10 \\
\tableline
\end{tabular}
\tablenotetext{a}{$-2\ln(L_{0}/L)$, 
where $L$ and $L_{0}$ are the
 maximum likelihoods for the model with/without the source component,
 respectively.}
\tablenotetext{b}{Three point sources listed in the 1FGL source catalog
 in the vicinity of the SNR G8.7$-$0.1, which their positions were free
 in the optimization.}
\tablenotetext{c}{The values are obtained by using the spatial templates
 from the various extracted regions, where the regions were determined
 by changing a lower limit from 0 to 15\% of the peak emission.}
\tablenotetext{d}{A point source model which is added to the Uniform
 disk and point source model.}
\end{center}
\end{table}

\begin{table}
\begin{center}
\caption{Parameters of the models for the \emph {Fermi} LAT sources.\label{tab:model}}
\begin{tabular}{lccccccccc}
\tableline\tableline
 Model  & \kep\tablenotemark{a} &  $s_{\rm L}$\tablenotemark{b} & $p_{\rm b}$\tablenotemark{c} & $s_{\rm H}$\tablenotemark{d} & $B$ &
 $\nh$\tablenotemark{e} & $W_{p}$\tablenotemark{f} & $W_{e}$\tablenotemark{f} \\
   &    &  & (GeV~$c^{-1}$) & &  ($\mu$G) & (cm$^{-3}$) & ($10^{49}$~erg) & ($10^{49}$~erg) \\
\tableline
(a)~Pion ($\nh = 100\ \mathrm{cm}^{-3}$)  & 0.01  & 2.0 & 3 & 2.7 & 100 & 100 & 2.8 & 4.6~$\times$~$10^{-2}$ \\
(b)~Pion ($\nh = 1000\ \mathrm{cm}^{-3}$) & 0.01  & 2.0 & 3 & 2.7 & 400 & 1000 & 0.30 & 7.2~$\times$~$10^{-4}$ \\
(c)~Bremsstrahlung & 1 & 2.0 & 5 & 2.7 & 25 & 100  & 0.22 & 0.36  \\
(d)~Inverse Compton\tablenotemark{g} & 1 & 2.0 & 15 & 3.5 & 1 & 0.1 & 48 & 99 \\

\tableline
\end{tabular}
\tablenotetext{a}{The ratio of electron and proton distribution functions
at 1~GeV~$c^{-1}$.}
\tablenotetext{b}{
The momentum distribution of particles
 is assumed to be a broken power-law, where the indices and the break
 momentum are identical for both accelerated protons and electrons.
$s_{\rm L}$ is the spectral index below the momentum break.}
\tablenotetext{c}{$p_{\rm b}$ is the momentum break for the particle distribution.}
\tablenotetext{d}{The spectral index for the broken power-law function
 above the momentum break.}
\tablenotetext{e}{Average hydrogen number density of the ambient medium.}
\tablenotetext{f}{The distance is assumed to be 4~kpc. The total energy
 is calculated for particles $>$~100~MeV~$c^{-1}$.}
\tablenotetext{g}{Seed photons for inverse Compton scattering of
 electrons include the cosmic microwave background,
 two infrared~($T_{\rm IR} = 37, 4.7 \times 10^2$~K, $U_{\rm IR} = 1.1,
 $0.23~eV~cm$^{-3}$, respectively), and
 two optical components~($T_{\rm opt} = 3.3 \times 10^3, 9.5 \times
 10^3$~K, $U_{\rm opt} = 1.2, 0.32$~eV~cm$^{-3}$, respectively) in the
 vicinity of G8.7$-$0.1, assuming a distance of 4~kpc.}

\end{center}
\end{table}

\end{document}